\documentclass[a4paper,11pt]{article}
\usepackage{enumitem}
\usepackage{hyperref}
\usepackage{booktabs}
 
\usepackage{authblk}
\usepackage{lettrine}
\usepackage{lmodern}

\hypersetup{
    colorlinks = true,
    linkcolor = Blue,
    citecolor = blue,
    urlcolor = blue
}

\usepackage{fontawesome5}

\makeatletter
\newcommand{\github}[1]{%
   \href{#1}{\faGithubSquare}%
}
\makeatother

\usepackage{cancel}
\usepackage{etoolbox}

\usepackage{fancyhdr}
\usepackage[dvipsnames]{xcolor}

\usepackage{amsmath,bm}
\usepackage{mathtools}
\hypersetup{colorlinks=True, citecolor=blue} 
\usepackage{amsfonts}
\usepackage{amssymb}
\usepackage{float}
\usepackage{bm}
\usepackage{graphicx}
\usepackage{color}
\usepackage{verbatim}
\usepackage{blindtext, multicol}
\usepackage{multirow}
\usepackage{epigraph}
\usepackage[dvipsnames]{xcolor}
\usepackage{subcaption}
\def\beq{\begin{equation}}
\def\eeq{\end{equation}}
\def\ber{\begin{eqnarray}}
\def\eer{\end{eqnarray}}
\def\benu{\begin{enumerate}}
\def\eenu{\end{enumerate}}

\def \blue {\color{blue}}

\def \Brown {\color{Brown}}

\def \dgray {\color{darkgray}}
\def \Blue {\color{Blue}}
\def \gray {\color{gray}}

\def\SSMR#1{{\color{red!25!black} #1}} 
\def\SSMB#1{{\color{blue!30!black} #1}}
\def\SSMG#1{{\color{green!30!black} #1}}
\def\SSMV#1{{\color{violet!80!black} #1}}

\usepackage{framed}
\colorlet{shadecolor}{CadetBlue!6}

\RequirePackage[top=20mm,bottom=22mm,left=20mm,right=20mm]{geometry}

\definecolor{headergray}{gray}{0.30}
\definecolor{rulegray}{gray}{0.72}

\fancypagestyle{articleheader}{
    \fancyhf{}

    \fancyhead[L]{%
        \footnotesize\color{Brown}\ttfamily
        Compact Objects in the Universe%
    }

    \fancyhead[R]{%
        \footnotesize\color{Brown}\ttfamily
        White Dwarfs, Neutron Stars and Black Holes%
    }

    \renewcommand{\headrule}{%
        \hbox to\headwidth{%
            \color{rulegray}\leaders\hrule height 1.5pt\hfill%
        }%
    }

    \fancyfoot[C]{\thepage}

}

\usepackage[sorting=none,style=numeric-comp,giveninits=true]{biblatex}
\usepackage[english]{babel}
\usepackage{csquotes}
\DeclareUnicodeCharacter{0301}{\'{i}}
\usepackage[nottoc]{tocbibind}

\usepackage{academicons}
\usepackage{fontawesome5}
\definecolor{orcidlogocol}{rgb}{0.65, 0.807, 0.223}
\newcommand{\orcid}[1]{$\,$\href{https://orcid.org/#1}{\textcolor{orcidlogocol}{\faOrcid}}}

\begin{document}
\vspace{-0.5in}
\title{\bf\huge \SSMB{Cosmic Corpses:~Dead stars as Factories \\of the most Extreme Objects in the Universe}}
\date{}

\author[1]{\SSMR{\large \bf Ankit C.~Khunt\,\orcid{0000-0002-5615-4645}}\thanks{Email ID: \texttt{{\blue ack.gravity@gmail.com}}}}
\author[2]{\SSMR{\large \bf Swagat S.~Mishra\,\orcid{0000-0003-4057-145X}}\thanks{Email ID: \texttt{{\blue swagatmishra@ibs.re.kr}}}}
\affil[1]{{\small Department of Physics, Sardar Patel University, Vallabh Vidyanagar 388120, India.}}
 \vspace{0.08in}
\affil[2]{{\small Cosmology, Gravity \& Astroparticle Physics Group, Centre for Theoretical Physics of the Universe (CTPU-CGA), Institute for Basic Science (IBS), 55 Expo-ro, Daejeon 34126, South Korea.}}
\maketitle
\thispagestyle{articleheader}
\vspace{-0.65in}

\begin{center}
{\gray
\noindent\rule{0.99\textwidth}{0.8pt}
}
\end{center}
\vspace{-0.3in}
\begin{shaded}
\noindent{\bf \Brown Prologue:}~\,The apparent serenity of the night sky conceals a long history of stellar evolution, in which gravity and the properties of matter determine the remarkable fates of stars. In this pedagogical account, we discuss white dwarfs, neutron stars, and black holes, explaining how quantum mechanics, nuclear interactions, and General Relativity govern their formation and distinctive properties. We also highlight the possibility of primordial black holes, which may have formed from rare large density fluctuations in the early Universe, and briefly discuss their potential role as constituents of dark matter.
\end{shaded}
\vspace{-0.35in}
\begin{center}
{\dgray
\noindent\rule{0.99\textwidth}{1.2pt}
}
\end{center}
\begin{multicols}{2}
\lettrine[lines=3]{O}n a clear night, the stars appear calm, beautiful, and almost eternal. They seem to belong to the ever-lasting architecture of the heavens, shining with an unchanging serenity across the years. Yet this impression is deceptive. Stars, too, have life histories: they are born, they evolve, and in time they die~\cite{kippenhahn_2012}. And often, it is in their final moments that they reveal some of the most dramatic phenomena in the Universe.

Even as we speak, somewhere in the Milky Way, stars are approaching the end of their lives. Somewhere in a distant part of our Galaxy, a massive star may have collapsed in the recent past, its final light still making its long journey towards us. The Milky Way is roughly $100,000$ light years across, so an event that appears in our sky today may in fact have occurred tens of thousands of years ago.

The speed of light, almost instantaneous on human scales, is nevertheless finite; and on cosmic scales this gives astronomy a memory. We see the heavens not as they are now, but as they were when their light began its journey. Within our own Galaxy this may carry us tens of thousands of years into the past, while light from the most distant galaxies we observe takes us back more than thirteen billion years. In this sense, a telescope looks not only across space, but also into time. To understand how stars meet their end, however, we must first understand how they live.

A star generally begins its life within a vast cloud of gas and dust scattered through the interstellar medium. Under its own gravity, the cloud slowly contracts, becoming denser and hotter towards the centre. As the contraction proceeds, the central temperature and density eventually become high enough for hydrogen fusion to commence. Nuclear fire is kindled in the core, hydrogen starts to fuse into helium, and {\Brown\bf \texttt{A STAR IS BORN}}. Once fusion is established, the star settles into a long-lived state of `hydrostatic equilibrium', in which the pressure of its hot interior supports the enormous weight of the layers above. The energy released in the core continually replenishes this thermal support, allowing the star to shine with remarkable steadiness for millions, or even billions, of years.

Yet the reservoir of hydrogen at the stellar core is finite. As this fuel is gradually depleted, the star can no longer maintain the same internal structure that sustained it through the greater part of its life. The core contracts under gravity and grows hotter; the surrounding layers respond, and the quiet stability of the main sequence gives way to a new stage in the stellar evolution. What follows is governed, above all, by the mass with which the star began its life. It is this that largely determines whether the star ends by leaving behind a white dwarf, whether its core collapses in a {\em supernova} to form a neutron star, or whether gravity carries the collapse still further, giving birth to a {\bf \texttt{BLACK HOLE}}\,---\,an extreme spacetime configuration from which even light cannot escape.

Yet black holes need not all be the relics of dead stars. Some may have an origin far more ancient, arising from exceptionally dense regions in the early Universe, long before the first stars had begun to shine~\cite{Hawking:1971ei}. We shall return to these `{\em primeval}', or `{\em primordial}', black holes later. For the moment, let us remain with the stars and follow what becomes of them when their long lives draw to a close. 

White dwarfs, neutron stars, and black holes belong
to a remarkable family of astronomical bodies known
as {\em compact objects}. These extraordinary objects pack enormous amounts of mass into remarkably small volumes, creating some of the most extreme environments known in nature. They allow astronomers to study matter at nuclear densities, magnetic fields trillions of times stronger than Earth's, and gravity operating at its most powerful limits.

\begin{figure}[H]
    \centering
    \includegraphics[width=0.99\linewidth]{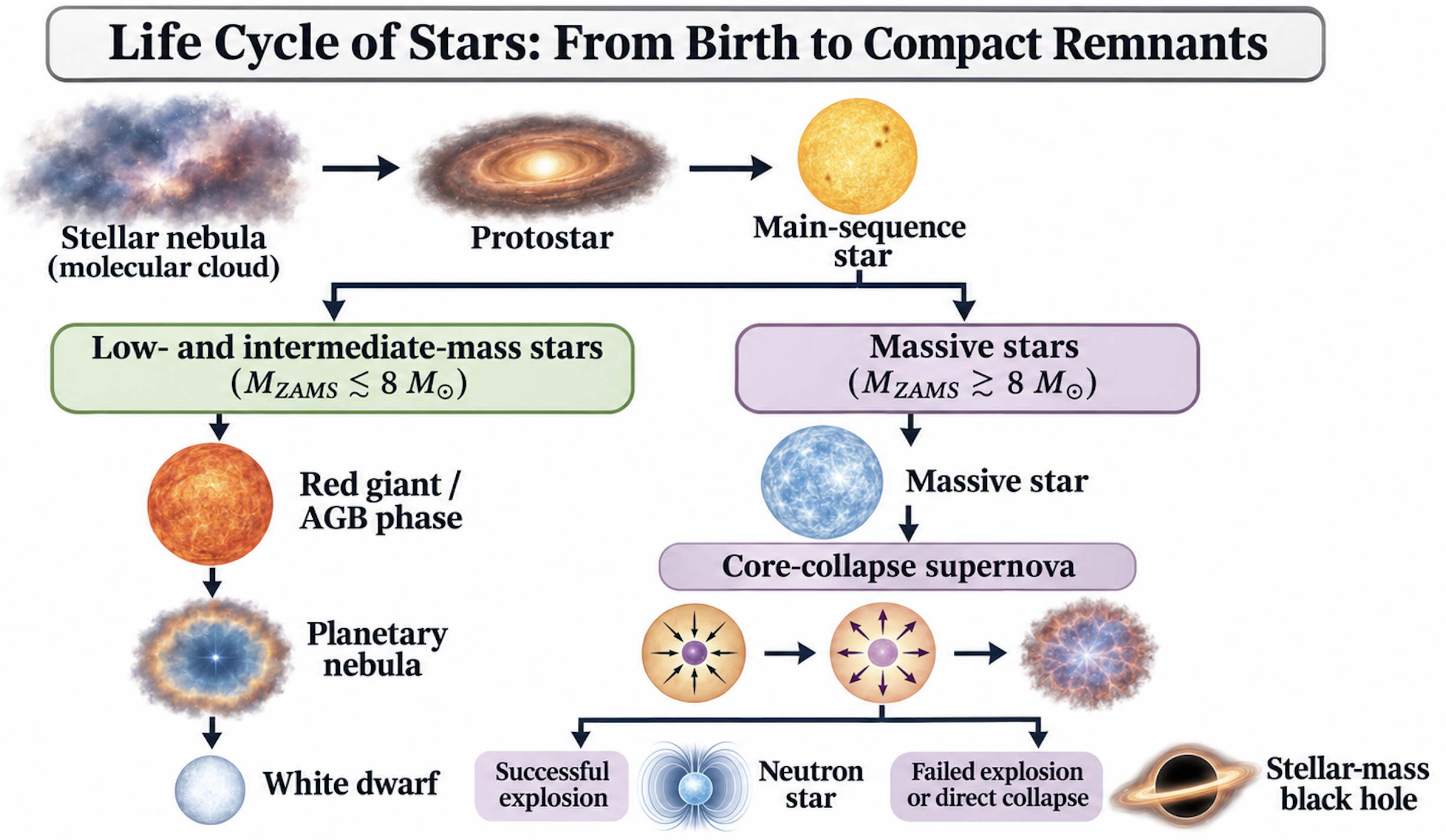}
    \caption{Schematic illustration of stellar evolution through low-mass and intermediate-mass and massive-star pathways, leading to white dwarfs, neutron stars, and stellar mass black holes through envelope loss, core-collapse supernovae, fallback, or direct gravitational collapse. {\gray[Author-assisted AI illustration]}.}
    \label{fig:Life_cycle}
\end{figure}

\vspace{-0.2in}

Compact objects are therefore not merely the ashes left behind by dying stars; they are places where nature presses matter, gravity, and quantum mechanics towards their limits. To see how such variety of distinct  dense objects can emerge from the lives of stars, let us now follow three stellar evolution paths, distinguished chiefly by mass, and watch how each approaches its final state, as illustrated in  Fig.~\ref{fig:Life_cycle}.

\subsection*{White Dwarfs: The Quiet Fate of Sun-like Stars}
For a star like our Sun, the end comes without a violent supernova. Once hydrogen in the core has been exhausted, the core contracts under its own gravity and heats up, while hydrogen continues to burn in a surrounding shell. The outer layers respond by expanding enormously, and the once familiar star enters its {\em \Brown red giant} phase. In time, the core becomes hot enough for helium itself to burn, producing predominantly carbon and oxygen. Our Sun will pass through these stages billions of years from now, growing so large that its outer atmosphere will extend to roughly the present orbit of the Earth.

\medskip

\begin{figure}[H]
\centering
\includegraphics[width=0.9\linewidth]{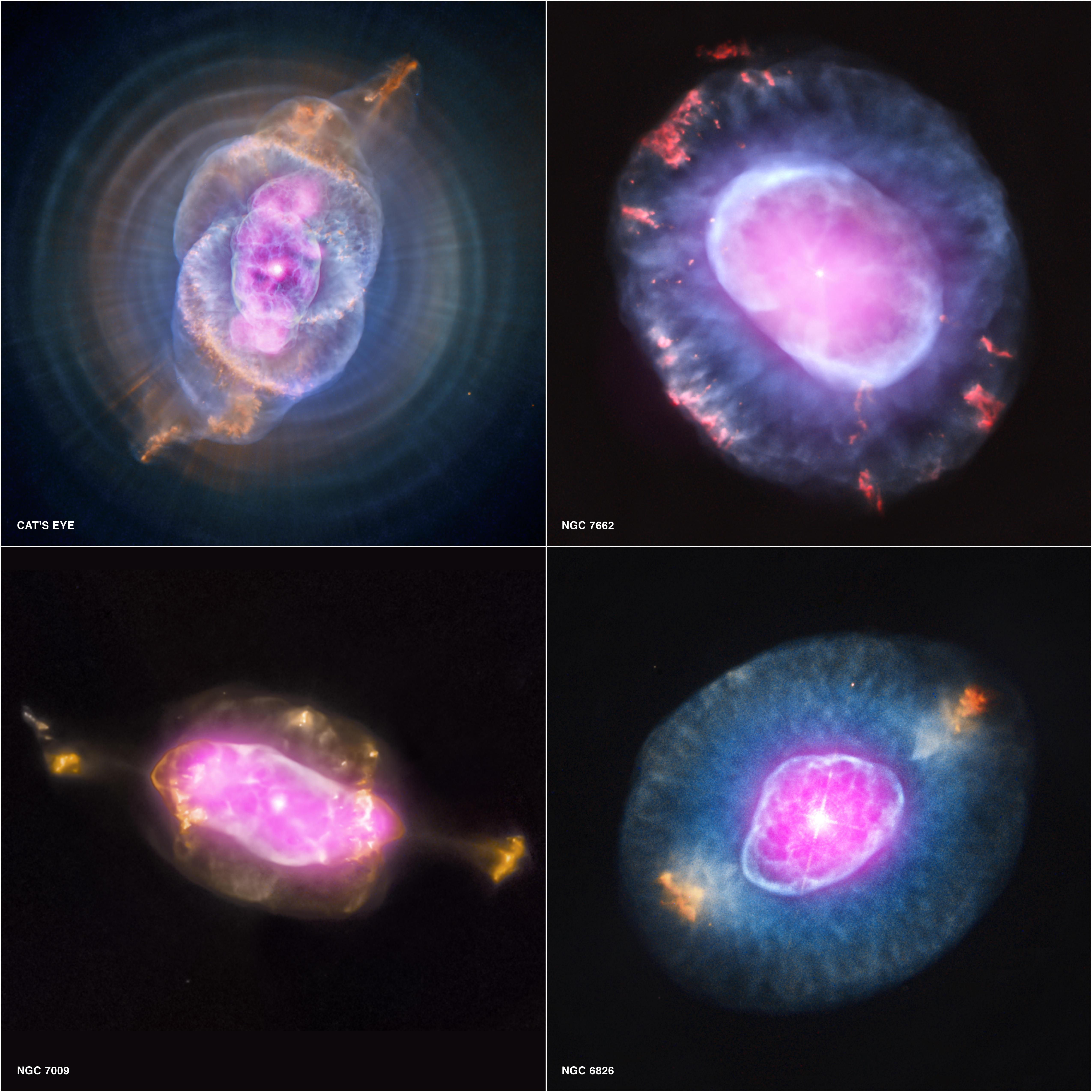}
\caption{Examples of four planetary nebulae, namely,  NGC 6543 (Cat's Eye Nebula), NGC 7662, NGC 7009 and NGC 6826, formed as  dying stars expelled their outer layers and exposed their hot central remnant. The ultraviolet radiation from the central star illuminates the surrounding shells of gas, offering a glimpse of the evolutionary path that eventually leaves behind a white dwarf.  {\gray  [Source: NASA, ESA, HEIC, and The Hubble Heritage Team (STScI/AURA);~\href{https://chandra.harvard.edu/photo/2012/pne/}{\texttt{Chandra X-ray Observatory}}]}.}
\label{fig:catseye}
\end{figure}

This, however, cannot continue indefinitely. After the central helium is exhausted, nuclear burning survives for a while in shells around the inert carbon--oxygen core, while the outer envelope becomes increasingly extended and unstable. Strong stellar winds and pulsations gradually cast much of this material into space. For a brief astronomical interval, the exposed hot core illuminates the departing gas, producing one of the most beautiful sights in the sky: a {\em planetary nebula}. The name is a historical accident, inherited from early telescopes in which these glowing clouds appeared vaguely planet-like. They have, of course, nothing to do with planets. Examples of such stellar farewells, including the Cat's Eye Nebula, are shown in Fig.~\ref{fig:catseye}.

What remains at the centre is small, hot, and extraordinarily dense. Its nuclear fuel is essentially exhausted, yet gravity does not succeed in crushing it further. At this point, a distinctly quantum mechanical form of support\,--\,the electron degeneracy pressure which is ultimately rooted in the {\em Pauli exclusion principle}\,--\,takes over.

Electrons are fermions, and the Pauli exclusion principle forbids them from all occupying the same quantum state~\cite{pauli_1925}. As matter is compressed into an ever smaller volume, the available low momentum states become filled and additional electrons are forced into states of progressively higher momentum. The resulting resistance to compression manifests itself macroscopically as {\em electron degeneracy pressure}. Unlike ordinary thermal pressure, it does not require a continuing source of nuclear energy and would persist even as the star became cold. Quantum mechanics, a theory born to describe the microscopic world of atoms and particles, thus becomes responsible for supporting an object of astronomical dimensions.

The remnant supported in this way is a \SSMV{\textbf{\texttt{WHITE DWARF}}}~\cite{shapiro_1983}. A typical white dwarf contains around $0.6\,M_\odot$, within a sphere comparable in size to the Earth. Its mean density can reach about a million grams per cubic centimetre. At such densities, a teaspoonful of matter would have a mass of several tonnes. The familiar scale of a planet and the mass of a star have, in a white dwarf, been brought together.

There is, however, a curious consequence of electron degeneracy: the more massive a white dwarf becomes, the smaller its radius. As the stellar mass increases, the gravitational attraction grows more rapidly than the degeneracy pressure can compensate at the original radius. The star must therefore contract, packing electrons into a smaller volume and, through the Pauli exclusion principle, forcing them to occupy states of higher momentum. The resulting increase in degeneracy pressure eventually restores equilibrium, but at a smaller radius. For a cold white dwarf supported by non-relativistic degenerate electrons, this behaviour is described by the approximate mass-radius relation $\bm{R\propto M^{-1/3}}$, where $R$ and $M$ denote the radius and mass of the star, respectively, assuming a fixed composition. This peculiar relation cannot continue indefinitely. At sufficiently high densities, the electrons become relativistic, changing the manner in which degeneracy pressure responds to further compression.

It was Subrahmanyan Chandrasekhar who showed that this leads to a limiting mass of $\bm{M_{\rm Ch}\simeq 1.4\,M_\odot}$ (where $M_\odot\simeq 2\times 10^{30}\,\mathrm{kg}$ denotes the solar mass), 
with the precise value depending somewhat on the composition and physical state of the star~\cite{chandrasekhar_1931}. Beyond the {\em \textbf{Chandrasekhar limit}}, electron degeneracy pressure can no longer provide a stable white dwarf configuration. This is one of the profound lessons of compact objects: even quantum pressure has limits to the weight it can support. Depending on the composition and evolutionary history, a white dwarf approaching the Chandrasekhar limit may undergo either thermonuclear disruption, as in a Type Ia supernova, or accretion-induced collapse to form a neutron star.

A white dwarf, notably, is not born cold. The newly exposed remnant is extremely hot and continues to shine, not through fresh nuclear reactions, but by slowly releasing the thermal energy stored within it. Over immense stretches of time it cools and fades, passing from brilliant white to ever dimmer states. Its long future is therefore less like the continued burning of a star and more like the fading glow of an ember after the fire itself has ceased. Given enough time, a white dwarf should ultimately become a cold and nearly invisible {\em \textbf{black dwarf}}. But the Universe is only about $13.8$ billion years old, far too young for any white dwarf to have completed this extraordinarily slow cooling process. Black dwarfs therefore belong, for now, to the distant future of the cosmos rather than to the catalogue of objects we can observe.

For stars such as the Sun, this is the quiet destiny that awaits, with their outer layers  returned to interstellar space, and their central remnant left behind as an Earth-sized monument to the quantum nature of matter. But electron degeneracy can carry only a finite burden! In more massive stellar cores beyond the Chandrasekhar limit, even this first quantum defence against gravity fails, and nature is driven towards a still denser form of matter. 

\subsection*{Neutron Stars: When Nuclear Matter Holds Back Gravity}

Stars born with eight or more times the mass of our Sun follow a rather different evolutionary course. Their greater masses allow them to reach temperatures at which successively heavier elements can undergo nuclear fusion. Hydrogen gives way to helium, followed by carbon, neon, oxygen and silicon, as the star progresses through increasingly brief stages of nuclear burning. These massive stars live not for billions of years, but for only millions or tens of millions of years. In many of them, this succession of nuclear reactions eventually produces a core rich in iron and neighbouring nuclei, collectively known as the {\em iron-group elements}. Here, however, the familiar nuclear process reaches its limit: further fusion of these nuclei can no longer release the energy needed to support the core from collapsing under gravity.

As the iron core grows, it approaches a critical state beyond which it can no longer remain stable. Electrons are captured by protons, while energetic photons begin to break apart heavy nuclei, further undermining the support against gravity. The iron-rich core collapses under its own weight in a fraction of a second. Matter is driven towards nuclear densities, and electrons combine with protons to produce neutrons, and ghostly particles called  {\em neutrinos}, schematically, through the reaction
$$
\mathrm{p}+e^{-}\longrightarrow \mathrm{n}+\nu_e\,.
$$
This process, known as {\em neutronisation}, changes the very composition of matter in the collapsing core.

The ensuing events are among the most violent ones in stellar evolution. As the inner core approaches nuclear density, the resistance of nuclear matter to further compression rises sharply, bringing its collapse to a halt and producing a rebound. A powerful shock wave is launched outwards, but its progress is soon impeded by the surrounding stellar material, prompting it to stall. Meanwhile, the newly formed, extremely hot neutron-rich stellar core releases an enormous amount of energy, predominantly in the form of neutrinos. A fraction of this neutrino energy, aided by the turbulent motion of matter behind the stalled shock, can revive the explosion. Consequently, the outer layers of the star are hurled into space in a so-called {\em core-collapse supernova}~\cite{bethe_1990}, a phenomenon whose connection with neutron stars was anticipated by Baade and Zwicky in the 1930s~\cite{baade_1934}.

\begin{figure}[H]
    \centering
    \includegraphics[width=1.0\linewidth]{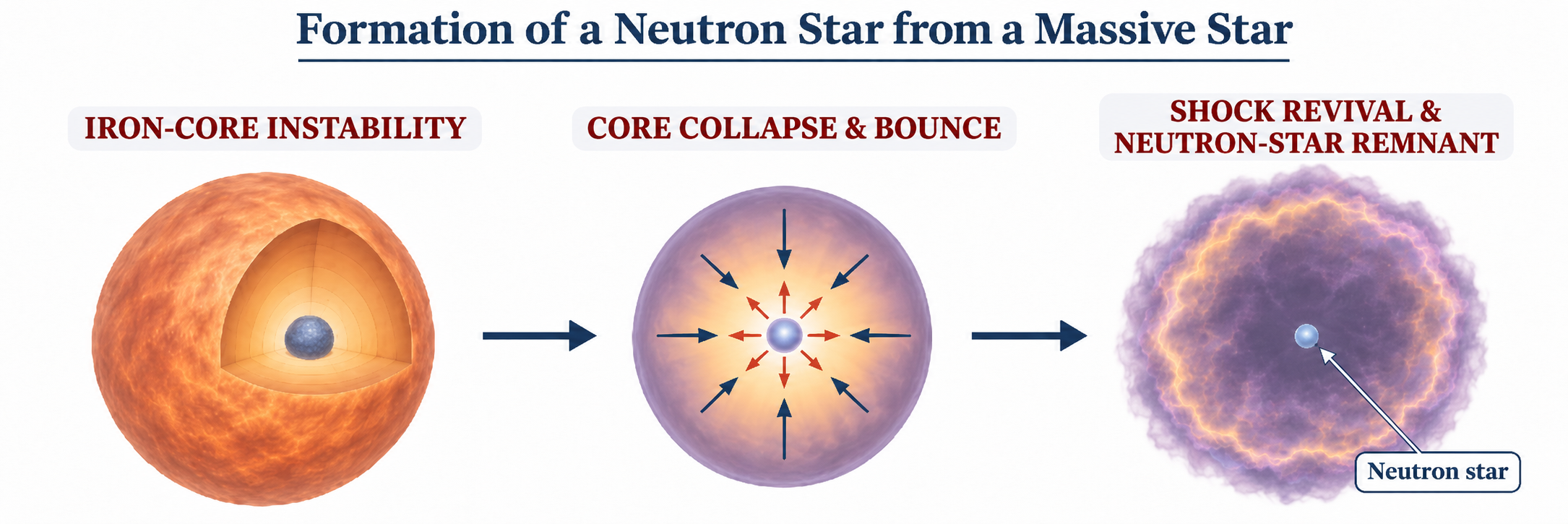}
    \caption{Schematic illustration of neutron-star formation through iron-core instability, gravitational collapse and core bounce, followed by shock revival and the formation of a neutron-star remnant {\gray [Author-assisted AI illustration].}}
    \label{fig:NS_structure}
\end{figure}

For a brief interval, the dying star may shine with the radiance of billions of ordinary stars. But beneath the expanding debris, a very different object has taken shape. The enormous stellar core has been compressed into a neutron-rich sphere only a few tens of kilometres across. This dense remnant, provided the collapsed core is not too massive to be supported, is a \SSMG{\textbf{\texttt{\textbf{NEUTRON STAR}}}}, see Fig.~\ref{fig:NS_structure}.

A typical neutron star contains about $1.4$ solar masses within a radius of merely $10$--$14$ kilometres. Its interior reaches densities of order $10^{14}$--$10^{15}\,\mathrm{g\,cm^{-3}}$, comparable to, and in its deepest regions often several times greater than, the density of matter within atomic nuclei. The familiar atomic structure of ordinary matter has largely disappeared. Beneath a crust containing nuclei and electrons lies matter overwhelmingly rich in neutrons, together with smaller populations of protons, electrons and other particles. It is tempting to picture the entire object as an enormous atomic nucleus, although the actual structure is considerably more intricate.

What prevents such an object from collapsing still further? The answer lies once again in quantum mechanics, but now the strong nuclear interaction also assumes a central role. Neutrons, like electrons, are fermions and obey the Pauli exclusion principle. Their degeneracy pressure, together with the pressure arising from interactions among densely packed nucleons, can sustain the star against its immense self-gravity. The equilibrium is governed not merely by Newtonian gravity, but by Einstein's General Theory of Relativity, whose effects are indispensable at such enormous compactness. Thus, the behaviour of matter on nuclear scales determines the fate of an object more massive than our Sun.

The birth of a neutron star also leaves its imprint upon its rotation. As the stellar core contracts, conservation of angular momentum can make it spin extraordinarily rapidly, much as a rotating skater spins faster upon drawing in their arms. Some neutron stars complete hundreds of revolutions every second. Their magnetic fields, inherited and greatly amplified during stellar evolution and collapse, can also be immensely strong. Radiation emitted from regions associated with the magnetic poles may sweep across space as the star rotates. Whenever such a beam crosses the Earth, our instruments register a regular succession of pulses, rather like the flashes from a distant lighthouse. These objects are known as {\em \textbf{pulsars}}. Their discovery by Jocelyn Bell Burnell in 1967 opened an entirely new chapter in observational astronomy~\cite{Hewish:1968bj}. The extraordinary regularity of some pulsars makes them natural cosmic clocks, capable of testing gravity with remarkable precision.

There are neutron stars whose magnetic fields reach still greater strengths. Known as {\em magnetars}, they may possess surface magnetic fields of $10^{14}$--$10^{15}$ gauss, hundreds of trillions of times stronger than the magnetic field of the Earth~\cite{Duncan:1992hi}. Such fields can store enormous quantities of magnetic energy, exert immense stresses upon the stellar crust, and power sudden outbursts of X-rays and gamma rays. In these objects, magnetic phenomena that are relatively modest in terrestrial laboratories acquire an almost unimaginable scale.

Yet perhaps the deepest mystery of neutron stars lies not in their rotation or magnetic fields, but in the state of the nuclear  matter hidden within them. At densities exceeding those of atomic nuclei,
the properties of matter are governed by the complex
many-body behaviour of the strong nuclear interaction,
whose attractive and repulsive contributions together
determine the pressure of matter under these extreme
conditions. The relation between pressure and density, known as the {\em equation of state}, remains theoretically uncertain in this regime. Does the inner core consist chiefly of neutrons and protons, or might it contain hyperons, deconfined quarks, or other unfamiliar phases of matter? Different possibilities lead to different predictions for the masses, radii and maximum stability of neutron stars. Observations have therefore become an important means of investigating physics that cannot yet be reproduced directly in terrestrial experiments.

An important contribution has come from NASA's {\em Neutron Star Interior Composition Explorer} (NICER), aboard the International Space Station. By studying X-ray pulsations from hot regions on neutron-star surfaces, and accounting for the bending of light in their strong gravitational fields, NICER observations constrain the masses and radii of these objects~\cite{Miller:2019cac}. Such measurements offer increasingly stringent tests of competing descriptions of nuclear matter at extreme densities.

\begin{figure}[H]
    \centering
    \includegraphics[width=1.0\linewidth]{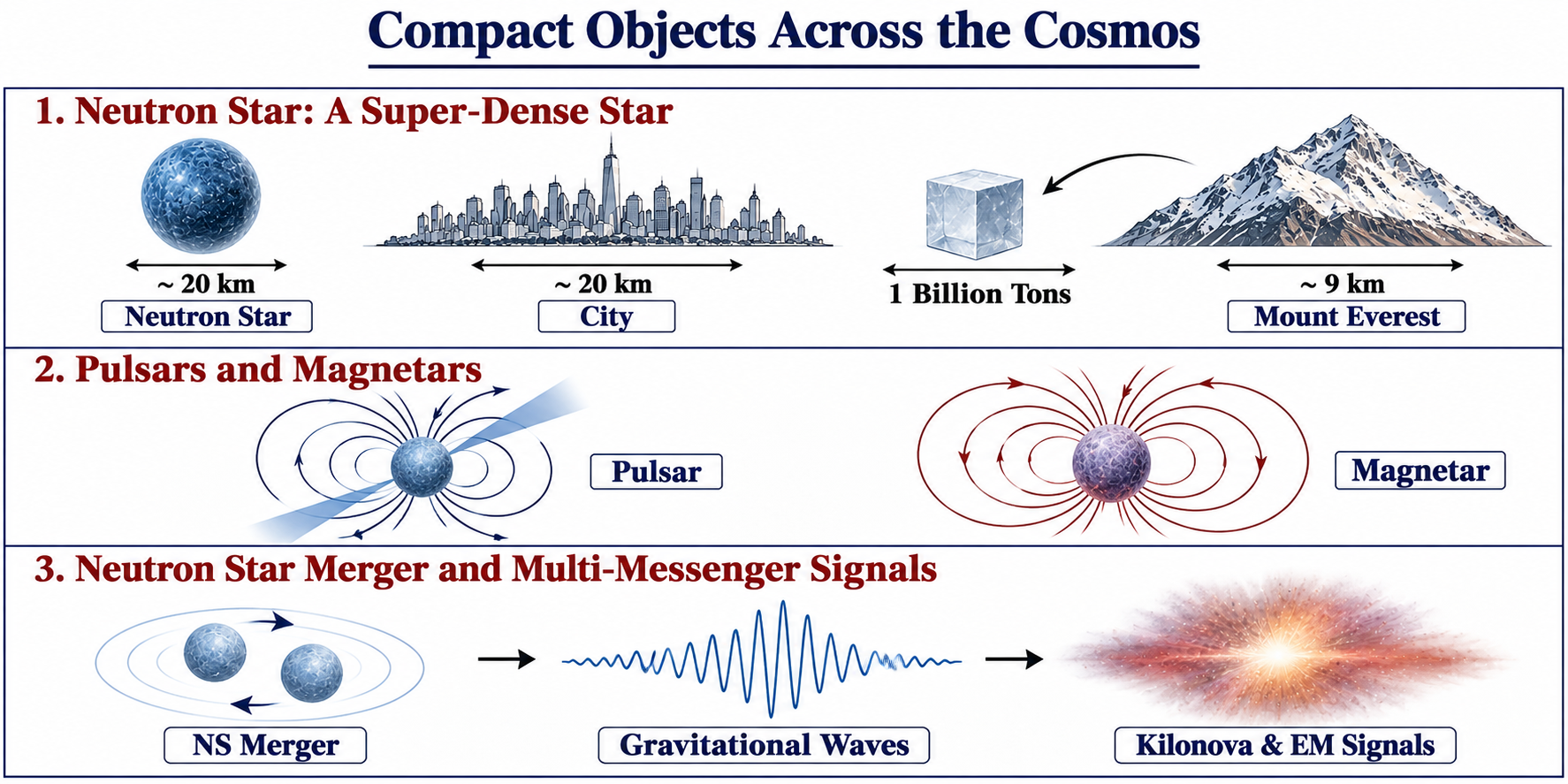}
    \caption{Illustration of neutron star compactness and extreme density, highlighting pulsars, magnetars, and neutron star mergers as sources of gravitational waves and electromagnetic signals. {\gray [Author-assisted AI  illustration].}}
    \label{fig:NS}
\end{figure}

Another significant discovery came in August 2017, when the LIGO and Virgo observatories detected gravitational waves from the merger of two neutron stars, an event designated \textbf{\texttt{GW170817}}~\cite{LIGOScientific:2017vwq}. Unlike the earlier detections of merging black holes, this collision was also observed through electromagnetic radiation, opening a new era of {\em multi-messenger astronomy}, see Fig.~\ref{fig:NS}.

The merger produced a luminous transient known as a {\em kilonova}, whose radiation provided compelling evidence for the synthesis of heavy elements through rapid neutron capture, the so-called {\em r-process}. Elements such as gold, strontium and platinum are thought to be produced in such neutron-rich environments~\cite{Watson:2019xjv}. Some of the heavy elements found on Earth may thus have been forged in collisions of neutron stars long before the formation of our Solar System.

For all their remarkable resistance to compression, neutron stars cannot support an arbitrarily large mass. General Relativity, together with the equation of state of dense matter, imposes an upper limit on the mass of a stable, non-rotating neutron star, commonly associated with the work of Tolman, Oppenheimer and Volkoff~\cite{oppenheimer_1939}. The precise limit remains uncertain because the behaviour of matter at the highest densities is not yet fully understood. 
We know, however, that neutron stars with masses around twice that of the Sun exist, placing important constraints upon this limit. Beyond the maximum stable mass, no known form of pressure can prevent continued gravitational collapse. Matter has reached the end of its resistance, and the way is opened to an object of a fundamentally different character: {\bf black hole}, to which we turn our attention next.

\subsection*{Black Holes: When Gravity Overcomes Every Resistance}

With neutron stars, we encountered the last familiar form of stellar matter capable of resisting gravitational collapse. Beyond this limit, however, something more profound can occur than the formation of an even denser star. The continued collapse of matter can produce a region of spacetime from which no signal, not even light, can reach the distant Universe. Such a region is what we call a \textbf{\texttt{BLACK HOLE}}~\cite{schwarzschild_1916}.

The boundary surrounding this region is known as the {\em event horizon}. Unlike the surface of a star or a planet, it is not a material boundary that one could touch, but a boundary in the causal structure of spacetime itself. It is sometimes said that gravity becomes so strong that the escape velocity exceeds the speed of light. While this offers a useful intuition, the deeper explanation lies in Einstein's General Theory of Relativity. Within the event horizon, the geometry of spacetime is such that even outward-directed light cannot find a future path leading back to the external Universe. The event horizon thus marks a boundary beyond which events can no longer communicate with distant observers.

For the simplest case of a non-rotating black hole, the radius of the event horizon is given by the {\em Schwarzschild radius},
\begin{equation*}
\bm{R_{\rm s}=\frac{2GM}{c^2}
\simeq 3\times \left(\frac{M}{M_{\odot}}\right)\,{\rm km}},
\end{equation*}
where $M$ is the black hole mass. Thus, an object containing several times the mass of our Sun can possess an event horizon only a few tens of kilometres across. Real astrophysical black holes are generally expected to rotate, giving their surrounding spacetime a richer geometry, but the essential property of the horizon remains the same.

It is worth dispelling a popular misconception. Black holes are often imagined as enormous cosmic vacuum cleaners, drawing everything around them towards inevitable destruction. In reality, their gravitational influence at sufficiently large distances is no different from that of any other object with the same mass. If, by some impossible transformation, our Sun were replaced by a black hole of precisely the same mass (with roughly $3$\,km radius), the Earth would continue along essentially the same orbit. What would change disastrously would be the absence of sunlight, not the gravitational motion of our planet.

For much of the twentieth century, black holes remained largely within the province of theoretical physics, arising as solutions to Einstein's equations~\cite{misner_1973}. Their existence in nature was far from universally accepted. Gradually, however, astronomical observations began to reveal objects whose behaviour was difficult to explain otherwise. In certain binary systems, gas drawn from an ordinary star towards an unseen compact companion becomes heated to enormous temperatures, producing intense X-rays before reaching the event horizon. Elsewhere, the motion of stars around apparently empty regions at the centres of galaxies points to concentrations of mass far beyond what ordinary stellar matter can explain. The evidence for black holes thus accumulated long before we could observe their immediate surroundings.

\begin{figure}[H]
    \centering
    \includegraphics[width=1.0\linewidth]{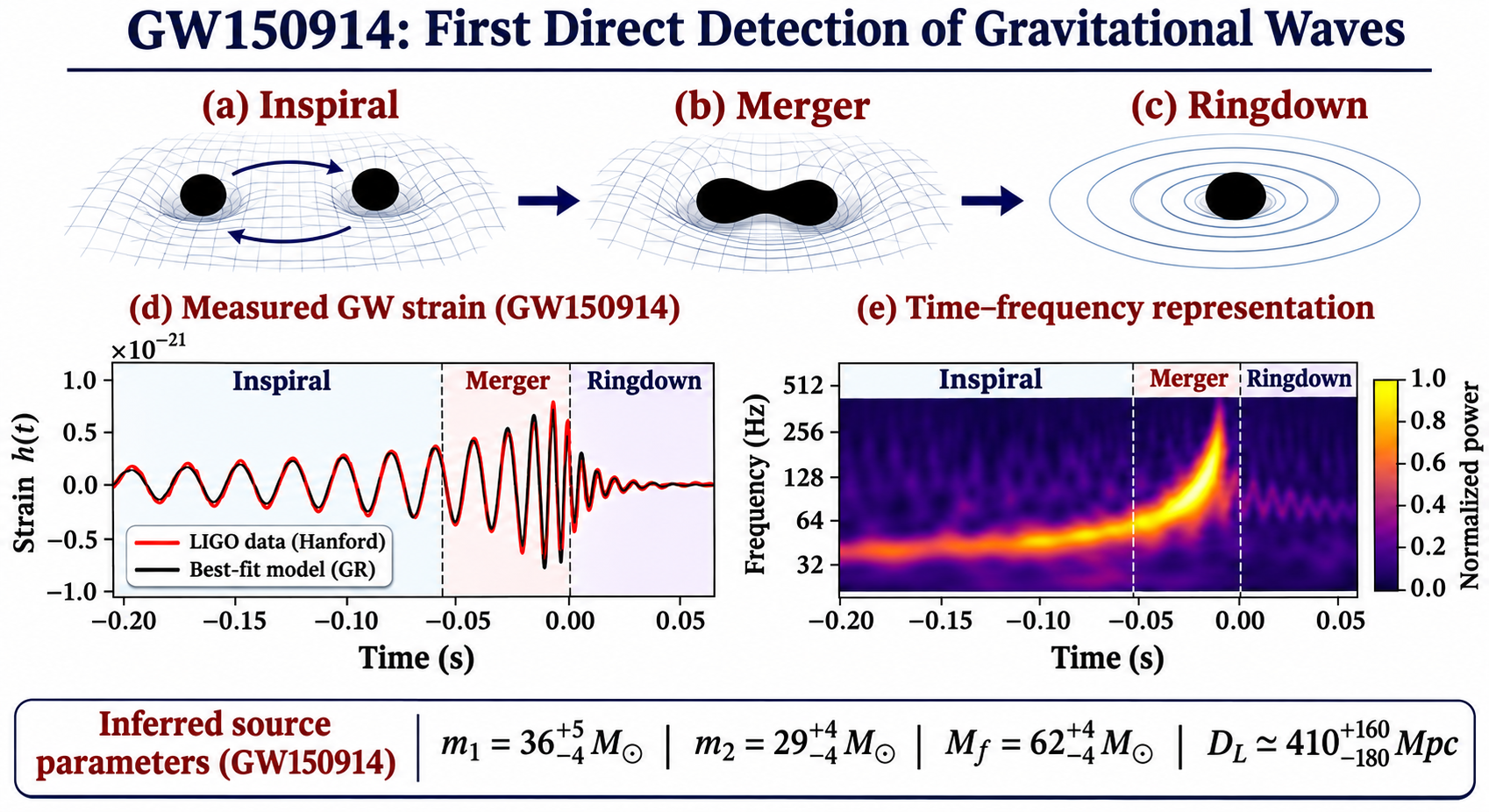}
    \caption{Schematic representation of the binary black-hole inspiral, merger, and ringdown phases, accompanied by the gravitational-wave strain and time–frequency evolution associated with GW150914. {\gray [Author-generated illustration. Waveform adapted from the \texttt{\blue LIGO Scientific Collaboration and Virgo Collaboration~\cite{abbott_2016}}.]}}
    \label{fig:BH_GW_1}
\end{figure}

A decisive advance came on 14 September 2015, when the two detectors of the Laser Interferometer Gravitational-wave Observatory (LIGO) recorded the first directly detected gravitational waves, produced by the merger of two black holes~\cite{abbott_2016}. As the black holes spiralled towards one another, they radiated energy through ripples in spacetime, drawing ever closer until they merged into a single rotating black hole. The observed signal carried the characteristic imprint of this process: a rising frequency and amplitude during the inspiral, followed by merger and the dying oscillations of the final object, known as the {\em ringdown}. For the first time, humanity had, metaphorically speaking,  `listened' to the final moments of two black holes through the vibrations of spacetime itself (Fig.~\ref{fig:BH_GW_1}).

\begin{figure}[H]
    \centering
    \includegraphics[width=1.0\linewidth]{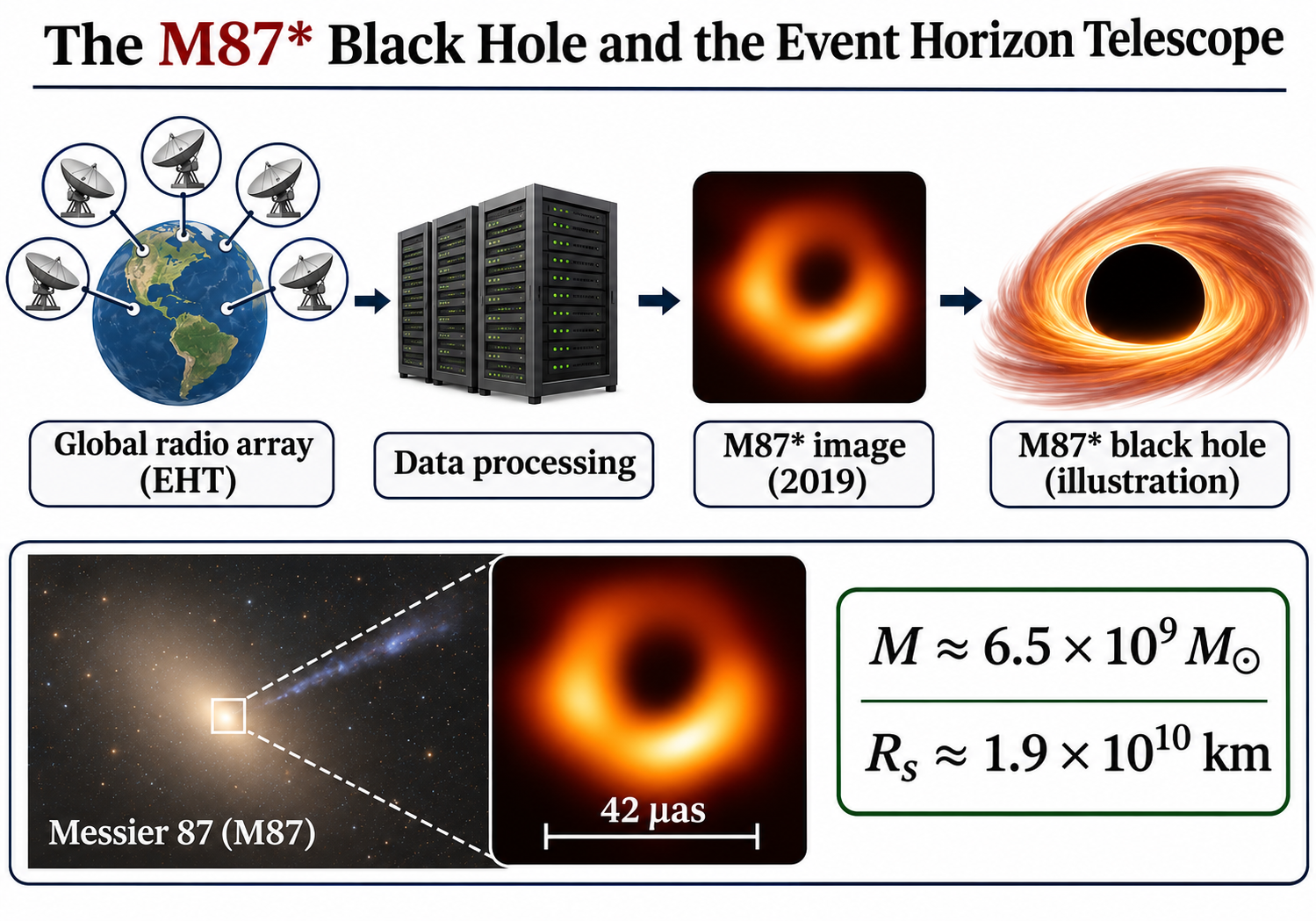}
   \caption{Illustration of the global Event Horizon Telescope network, data processing, and the reconstructed 2019 image of M87*, highlighting its characteristic emission ring and black-hole shadow. {\gray [Author-assisted AI illustration; black-hole image credit: \href{https://eventhorizontelescope.org/M87-one-year-later-proof-of-a-persistent-black-hole-shadow}{\texttt{Event Horizon Telescope Collaboration (2019)}}].}}
    \label{fig:m87_eht}
\end{figure}

\vspace{-0.2in}

A further milestone followed in April 2019, when the {\em Event Horizon Telescope} (EHT), a worldwide network of radio observatories operating together as an Earth-sized telescope, released the first image of the immediate surroundings of a black hole~\cite{collaboration_2019}. The object was M87*, the supermassive black hole at the centre of the galaxy Messier 87, containing some $6.5$ billion solar masses  (Fig.~\ref{fig:m87_eht}). The now familiar image revealed a bright ring of emission surrounding a dark central region, the {\em black hole shadow}~\cite{Synge:1966okc}. Strictly speaking, this was not an image of the event horizon itself, but of radiation from hot matter whose paths had been strongly bent by gravity, surrounding a dark region associated with the capture of light~\cite{Gralla:2019xty}. The observation offered a striking confirmation of the behaviour predicted by General Relativity in the neighbourhood of a black hole.

In 2022, the EHT collaboration revealed a similar image of Sagittarius~A$^*$, the supermassive black hole at the centre of our own Milky Way, whose mass is about four million times that of the Sun~\cite{eht_2022}. These observations, together with gravitational-wave detections and the motions of stars and accreting gas, have transformed black holes from objects once regarded as mathematical curiosities into an established part of the astronomical world.

Black holes exist across an astonishing range of masses. Those born from massive stars typically contain several to many tens of solar masses, while the giants found at the centres of galaxies may contain millions or billions of solar masses. The latter can grow by accreting surrounding matter and merging with other black holes, although the origin of their earliest seeds and the rapid growth of some of them remain open questions. Their enormous masses provide yet another reason to investigate the history and mechanism of black-hole formation.

Despite this impressive progress, some of the deepest conceptual questions remain beyond our present understanding. According to classical General Relativity, continued gravitational collapse can lead to a {\em spacetime singularity}~\cite{Raychaudhuri:1953yv,Penrose:1964wq,Hawking:1970zqf}. In idealised solutions, this is associated with divergent curvature and the breakdown of the classical description of spacetime. Rather than providing a complete account of what lies within, such predictions may be telling us that the theory has reached the boundary of its applicability. What truly happens deep inside a black hole remains unknown.

Quantum theory has made the puzzle still more intriguing. In the 1970s, Stephen Hawking showed that quantum fields in curved spacetime predict that black holes can emit thermal radiation, now known as {\em Hawking radiation}~\cite{Hawking:1974rv,Hawking:1975vcx}. This is not ordinary light escaping from within the event horizon, but a quantum phenomenon associated with the black-hole geometry. For astrophysical black holes, the predicted radiation is exceedingly weak and has not been directly observed. Nevertheless, its theoretical existence suggests that black holes can gradually lose mass, raising a profound question: what happens to the information carried by matter that falls into a black hole if the black hole ultimately evaporates? The resulting {\em black-hole information paradox}~\cite{Hawking:1976ra} stands at the meeting point of gravity, quantum theory and thermodynamics, and remains central to the search for a consistent theory of quantum gravity~\cite{Raju:2020smc}.

\bigskip

\begin{figure}[H]
    \centering
    \includegraphics[width=1.0\linewidth]{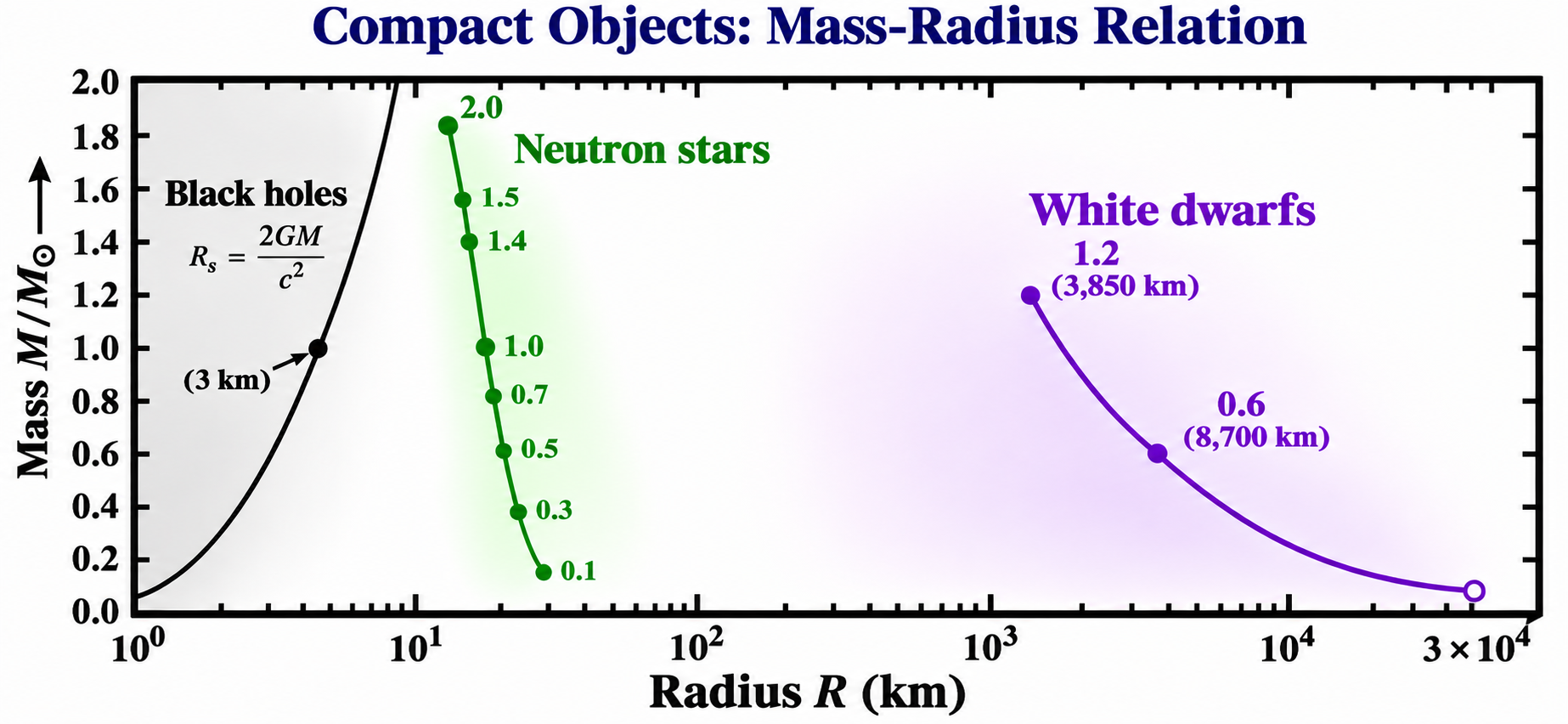}
   \caption{Schematic mass–radius relations of black holes (black), neutron stars (green), and white dwarfs (purple), illustrating the Schwarzschild horizon relation and characteristic stellar equilibrium sequences. Representative masses and radii highlight the compactness of neutron stars and the inverse mass–radius relationship of white dwarfs. {\gray[Author-generated illustration].}}
    \label{fig:Compactness}
\end{figure}

White dwarfs, neutron stars, and black holes thus illustrate a
remarkable progression in compactness, with stellar masses
confined to regions ranging from the size of a planet to only
a few kilometres. Their characteristic masses and radii
are compared in Fig.~\ref{fig:Compactness}, where the radius
of a black hole denotes the size of its event horizon rather than a
material surface.

\vspace{-0.1in}

We have, thus far, followed black holes from their beginnings in the collapse of massive stars to their place in some of the deepest questions of fundamental physics. Yet the existence of an event horizon tells us nothing of
the history of the matter that formed it. {\em Must every black
hole have once been a star}\,? To pursue this question, we
must turn from the deaths of massive stars to the earliest epoch in the history  of the Universe: the {\em hot Big Bang phase}.

\subsection*{Primordial Black Holes from the early Universe}

Black holes are commonly regarded as the graveyards of massive stars. Yet their defining feature is not the process by which they are formed, but rather the very structure of spacetime encompassing them. Broadly speaking, the presence of an event horizon, beyond which even light cannot escape to distant observers, is what characterises a black hole. Classical General Relativity also predicts singularities within familiar black-hole solutions, although their existence in a complete quantum theory of gravity remains uncertain. In principle, therefore, nature need not rely upon the death of a star to create a black hole. Any physical process capable of producing the necessary gravitational collapse would lead to a similar fate.

The possibility that black holes could have formed long before the appearance of first stars was explored in the early development of relativistic cosmology. Most notably,  Stephen Hawking in 1971 investigated the intriguing possibility that gravitational collapse in the early Universe could produce black holes of extremely small masses~\cite{Hawking:1971ei}. These hypothetical objects, now known as \textbf{\texttt{{\Blue PRIMORDIAL} BLACK HOLES}} (PBHs), differ from ordinary stellar black holes not in their fundamental gravitational nature, but in their origin. They might have formed through several distinct  processes in the early Universe, including the collapse of primordial density fluctuations, vacuum bubbles, or cosmic-string loops~\cite{Green:2020jor,Carr:2026hot,Escriva:2022duf}. Among these possibilities, the collapse of unusually large primordial density fluctuations remains one of the most extensively studied~\cite{Carr:1974nx,Green:2020jor}.

To understand this process, we must return to the hot, dense, and rapidly expanding Universe during its earliest history. Although the primordial Universe was remarkably smooth, matter and radiation were not distributed with perfect uniformity. Tiny variations in density were present, with some regions containing slightly more matter than their surroundings and others slightly less. With time, gravity amplified these differences, eventually giving rise to galaxies, clusters of galaxies, and the vast cosmic web that we observe today. But what if some of these primordial fluctuations had been much larger than those responsible for ordinary cosmic structures?

\begin{figure}[H]
    \centering
    \includegraphics[width=1.0\linewidth]{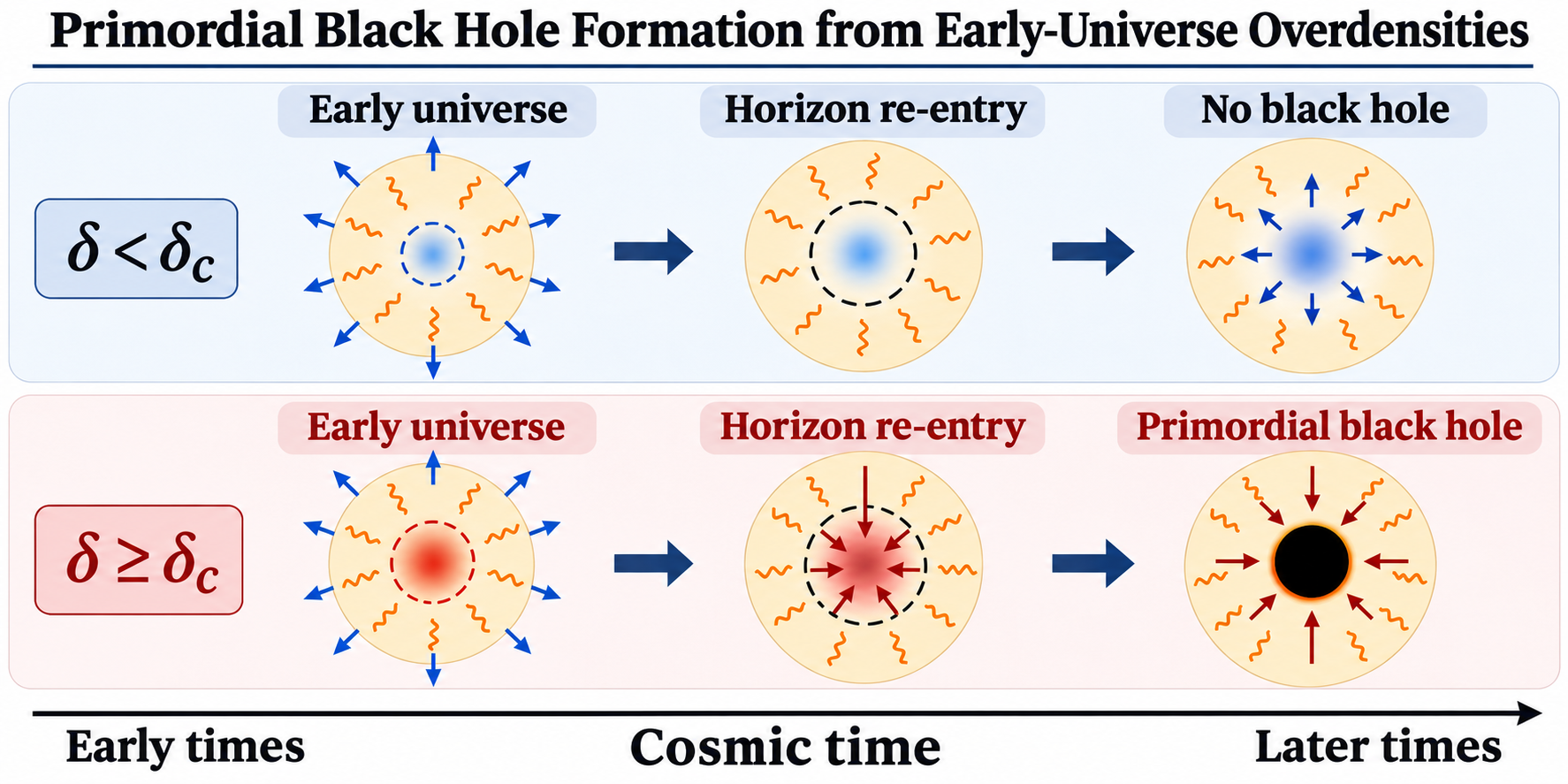}
    \caption{Schematic illustration of primordial black-hole formation from early-Universe density perturbations, contrasting subcritical overdensities that disperse after horizon re-entry with sufficiently large overdensities that undergo gravitational collapse. {\gray [Author-assisted AI illustration.]}}
    \label{fig:PBH1}
\end{figure}

In an expanding Universe, an overdense region need not always continue expanding with its surroundings. If its density contrast is sufficiently large when its physical scale becomes comparable to the Hubble radius, its gravitational attraction can overcome the pressure of the surrounding radiation. Such a region may cease to follow the general expansion and collapse under its own gravity, eventually forming a black hole. Most primordial fluctuations would be far too small for this to happen. It is the rare, exceptionally large fluctuations that matter, see Fig.~\ref{fig:PBH1}. Their abundance depends upon the statistical distribution of primordial perturbations: a greater variance generally makes large fluctuations more likely, while the precise number of black holes depends upon the rare tail of that distribution and the conditions required for collapse. The resulting PBH masses are close to the mass enclosed within the Hubble radius at the epoch of formation, allowing black holes far lighter than those produced by stellar evolution~\cite{Carr:1974nx}.

But where did these primordial fluctuations originate? One of the most compelling answers comes from {\em \textbf{cosmic inflation}}, a period of rapid accelerated expansion that preceded the hot Big Bang phase~\cite{Mishra:2024axb}. In its simplest realisation, inflation is driven by a scalar field, the {\em inflaton}, whose potential energy dominates the cosmic expansion. As the inflaton slowly rolls down its potential, quantum fluctuations of the field are stretched to astronomical scales, leaving behind small variations in the curvature and density of the Universe. These fluctuations provide the seeds of the cosmic structures we observe today.

\begin{figure}[H]
    \centering
    \includegraphics[width=1\linewidth]{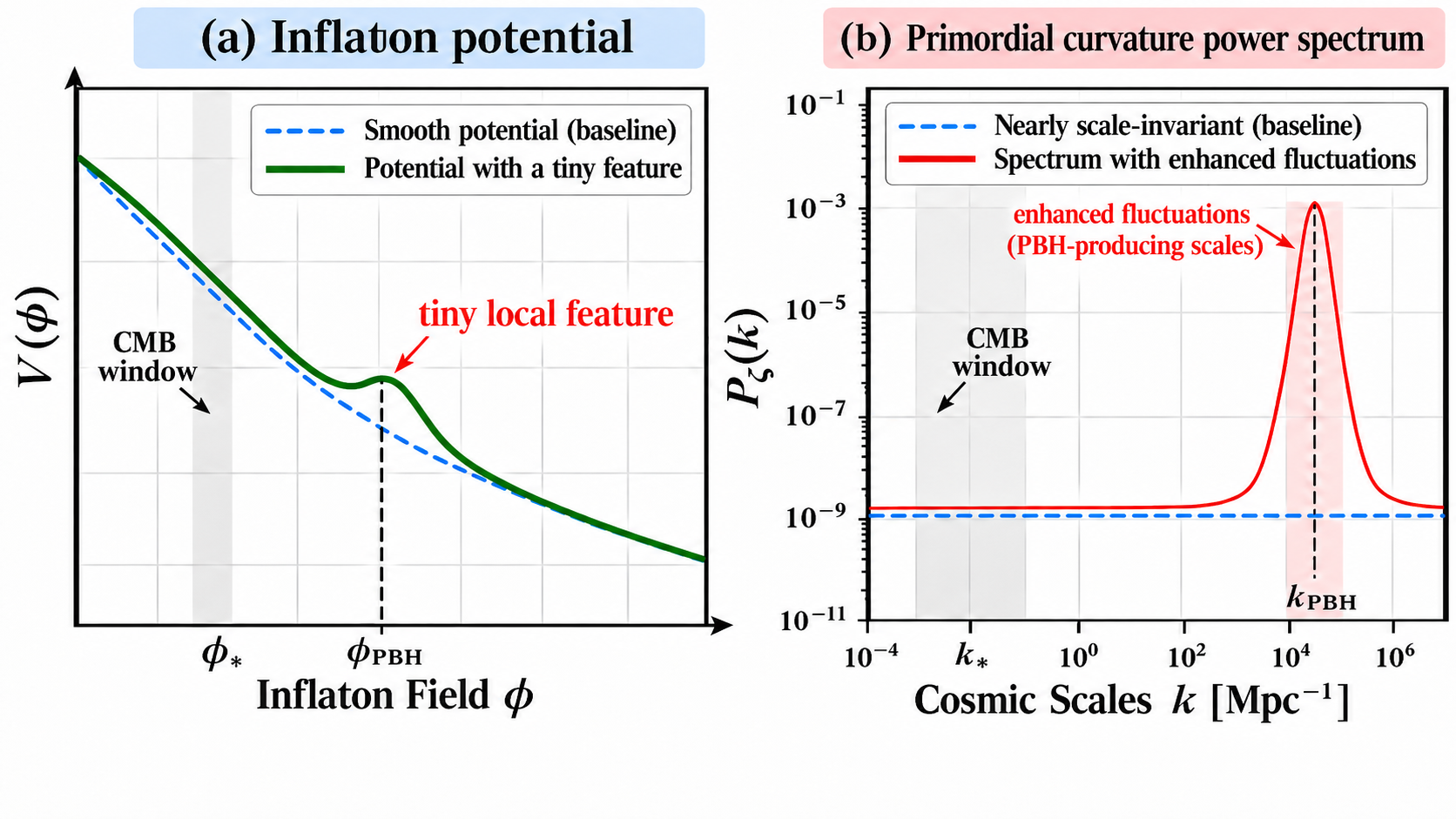}
    \vspace{-0.2in}
    \caption{Schematic illustration of the enhancement of primordial fluctuations during inflation. (a) A small local feature in the form of a speed-breaker 
    in the inflaton potential can temporarily alter the
    inflaton dynamics. (b) Under suitable conditions,
    this can produce a pronounced enhancement in the
    primordial curvature power spectrum on selected scales,
    increasing the probability of PBH formation. Inspired by the inflationary mechanism for PBH formation 
    discussed in Ref.~\cite{Mishra:2019pzq}. {\gray [Author-generated illustration.]}}
    \label{fig:PBH2}
\end{figure}

In conventional `slow-roll' inflation, the resulting perturbations are generally too small to produce an appreciable population of PBHs. However, even a tiny departure from the smooth evolution of the inflaton can change this picture~\cite{Ivanov:1994pa}. A small bump in its potential may act as a {\em speed-breaker}, temporarily slowing the field and considerably amplifying fluctuations on selected scales, as illustrated in Fig.~\ref{fig:PBH2}. Under suitable conditions, even a small dip can produce a similar enhancement. Such possibilities were explored in Ref.~\cite{Mishra:2019pzq} (and elsewhere), where the authors demonstrated that remarkably small local features in the inflaton potential can substantially enhance the primordial power spectrum and the subsequent probability of PBH formation. In this way, a seemingly insignificant feature in the physics of inflation may leave behind an entire population of black holes, born long before the first stars came into existence.

Primordial black holes are especially interesting because they may also help explain one of the enduring mysteries of cosmology: {\em \textbf{dark matter}}. Unlike ordinary matter, dark matter does not emit or absorb light in any appreciable way, yet its gravitational influence is evident throughout the Universe. PBHs, if sufficiently abundant and long-lived, could contribute to this unseen matter without requiring a new elementary particle~\cite{Chapline:1975ojl}. Their possible masses span an enormous range. Very light PBHs would lose mass through Hawking radiation and could have evaporated during cosmic history, while sufficiently massive ones may survive to the present epoch. Their abundance is constrained by several observations, including searches for radiation from evaporation, gravitational microlensing of distant stars, effects of accretion on the cosmic microwave background, and gravitational-wave measurements. Although many mass ranges are strongly restricted, certain possibilities, particularly around asteroid masses, remain viable for a substantial or even dominant contribution to dark matter~\cite{Green:2020jor,Carr:2026hot}.

No primordial black hole has yet been unambiguously identified. Nevertheless, the possibility of their existence offers a remarkable connection between the microscopic fluctuations of the earliest Universe and the gravitational phenomena we can investigate today. Stellar black holes reveal how massive stars meet their end; primordial black holes, if discovered, could carry traces of an era when neither stars nor galaxies had yet come into existence. They would offer us a new means of studying the infancy of the Universe, through the gravitational relics it may have left behind.

\end{multicols}

\medskip

\begin{shaded}
\noindent{\bf \Brown Epilogue:} \underline{\bf Coming Back to the Sky}\\[-0.05in]

We began beneath a clear night sky, where the stars seemed calm
and almost eternal. We return to it with a rather different
understanding. The end of a star's luminous life need not mean
the end of its influence upon the Universe. Its remnant may
endure for immense ages, revealing itself through pulses of
radiation, the motion of neighbouring stars, or ripples in
spacetime. And if primordial black holes exist, some of these
objects may be older than the first stars themselves.

\medskip

The study of compact objects has thus led us from familiar starlight to the deepest workings of matter and gravity. Quantum mechanics, nuclear forces, and the geometry of spacetime, ordinarily encountered as separate chapters of physics, come together in the lives of these objects. Through their observable properties, compact objects allow us to test the laws of nature under conditions that no terrestrial laboratory can reproduce, and perhaps to discover where our present understanding must be carried further.

\begin{center}
\texttt{\SSMB{This article is dedicated to the late\\[0.05in]
Prof.~Jayant~V.~Narlikar \& Prof.~Naresh~Dadhich.}}
\end{center}
\vspace{-0.08in}
\end{shaded}

\subsection*{Acknowledgement}
ACK acknowledges the Inter-University Centre for Astronomy and Astrophysics (IUCAA), Pune, for their hospitality during the preparation of this article  under  IUCAA's Visitor Programme.
SSM is supported by the Institute for Basic Science (IBS), Daejeon, as a Senior Researcher at CTPU-CGA under the project code, \texttt{IBS-R018-D3}. The authors acknowledge the use of large language models (LLMs), in particular \texttt{ChatGPT 5.6 Sol} (OpenAI) and \texttt{Gemini} (Google), in preparing some of the illustrative figures and in making stylistic edits to the text of this article.
\begin{center}
{\gray \noindent\rule{0.995\textwidth}{1.8pt}}
\end{center}

\vspace{0.00in}

\begin{multicols}{2}
\renewcommand*{\bibfont}{\footnotesize}
\setlength{\bibitemsep}{0.2\baselineskip}
\printbibliography
\end{multicols}
\end{document}